\documentclass[prl,showpacs,preprintnumbers,twocolumn]{revtex4}
\pdfoutput=1 
\usepackage{amsmath}
\usepackage{graphicx}

\newcommand{\out}[0]{\mathrm{out}}
\newcommand{\equ}[0]{\mathrm{eq}}
\newcommand{\ta}[0]{\mathrm{t}}
\newcommand{\ba}[0]{\mathrm{b}}
\newcommand{\inn}[0]{\mathrm{in}}
\newcommand{\dev}[0]{\mathrm{d}}
\newcommand{\diff}[0]{\mathrm{d}}
\newcommand{\e}[0]{\mathrm{e}}

\begin{document}
\title{On the energetics of information exchange}

\author{J. Hoppenau}
\author{A. Engel}

\affiliation{Carl von Ossietzky Universit\"at Oldenburg, 26111 Oldenburg, Germany}


\begin{abstract}
We consider the thermodynamic properties of systems in contact with an information source and focus on the consequences of energetic cost associated with the exchange of information. To this end we introduce the model of a {\em thermal tape} and derive a general bound for the efficiency of work extraction for systems in contact with such a tape. Depending on the perspective, the correlations between system and tape may either increase or reduce the efficiency of the device. We 
illustrate our general results with two exactly solvable models, one being an autonomous system, the other one involving measurement and feedback. We also define an {\em ideal tape limit} in which our findings reduce to known results. 
\end{abstract}

\maketitle

\section{Introduction}

\label{sec:into}
The advent of stochastic thermodynamics \cite{Seifert2012,Esposito2012a} revived the discussion about the link between statistical mechanics and information theory. Sharpening the Second Law of Thermodynamics into an equation, and unambiguously identifying information as a thermodynamic resource allowed an improved analysis of some generic problems at the borderline between statistical physics and information theory. Examples include thermodynamic efficiencies in the presence of information exchange \cite{Cao2009, Sagawa2010, Sagawa2012a,Deffner2013, Horowitz2011a, Granger2011, Abreu2012}, the formation of information carrying biomolecules in driven chemical systems\cite{Andrieux2008, Andrieux2009}, as well as the operation of models of Maxwell's demon \cite{Barato2013,   Abreu2011, Barato2013a, Mandal2012,  Vaikuntanathan2011, Horowitz2013}. 

In a paradigmatic setup a physical system interacts with a tape from which it may read and/or on which it may write information in form of bits. In most investigations so far, two idealized properties of such a tape have been taken for granted: first, the tape preserves the information forever i.e., its states do not change spontaneously; second, the reading and writing of information does not require energy. 

In the present paper we study the modifications arising when these idealizations are weakened. 
To this end we introduce the model of a {\em thermal tape}. The cells of such a tape are $n$-level systems with all levels having different energies $E(y)$. Moreover, the tape is connected with its own heat bath of temperature $T_\ta$. Changing the state of a cell of such a tape hence requires some energy; leaving a cell alone for a long time results in a relaxation to the equilibrium distribution of the tape. In this way the two features of an idealized tape mentioned above are replaced by somewhat more realistic assumptions. We first study energy and entropy balance of a general setup involving a thermal tape and then investigate two more specific examples. The first concerns an autonomous Maxwell's demon, the second deals with a systems with measurement and feedback. We investigate bounds on the thermodynamic efficiency of these models and discuss their relation to those obtained previously for ideal tapes. A convenient way to do so is to scale the energies of the tape states according to $E(y)=\epsilon(y) T_\ta$ and to perform the {\em ideal tape limit} defined by $T_\ta\to 0$.

\section{General model}

\begin{figure}
  \centering
  \includegraphics{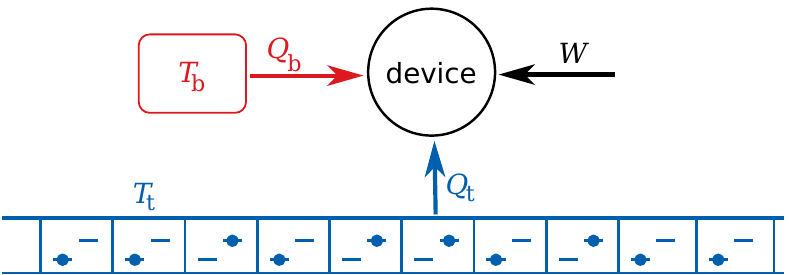}
  \caption{Setup: a device interacts with an heat bath at temperature $T_\ba$, a work reservoir, and a tape. The tape consists of identical $n$-level-systems. Before the interaction with the device, the tape is in equilibrium with a bath at temperature $T_\ta$.}
  \label{fig:fig2}
\end{figure}

We consider the setup depicted schematically in Fig.~\ref{fig:fig2}. A physical system called device exchanges heat $Q_\ba$ with a heat bath of temperature $T_\ba$ and work $W$ with a work source. In addition it interacts with an information carrying tape made of identical units which we model as $n$-level systems, $y=0,...,(n-1)$ with energies $E(y)$. The tape is coupled to its own heat bath of temperature $T_\ta$; we will assume $T_\ta<T_\ba$ throughout our analysis.

Before interacting with the device the cells of the tape are assumed to be in equilibrium, i.e. the probability to find a cell in state $y$ is given by 
\begin{equation}
  \label{eq:p_eq_tape}
  p_\ta^\equ(y) = \exp\left(\frac{F-E(y)}{T_\ta}\right),
\end{equation}
where 
\begin{equation}
 F = -T_\ta \ln \sum_{y=0}^{n-1} \exp\left(-\frac{E(y)}{T_\ta}\right).
\end{equation} 
Throughout this paper we set Boltzmann's constant  $k_b \equiv 1$.
Each cell interacts with the device for a given time interval $\tau$ during which it is decoupled from the tape bath. In the course of the interaction the system may change the state of the cell at the cost of supplying or receiving the suitable amount of energy. At the end of  the interaction period the cell is decoupled from the device and brought into contact with the tape bath again. It then starts to relax back to its equilibrium distribution. After interacting with a sufficient number of cells the device will in general approach a stationary state, in which its probability distribution at the end of the interaction interval coincides with the one at the beginning. We will  concentrate on this stationary regime.

Let us denote by $p^\out_\ta(y)$ the probability for a cell to be in state $y$ after the interaction with the device. Clearly, $p^\out_\ta(y)\neq p^\equ_\ta(y)$ in general.
For the average energy $Q_t$ exchanged between device and tape we then have
\begin{equation}\label{eq:defQt}
 Q_t=\sum_y E(y)[p^\equ_\ta(y)-p^\out_\ta(y)].
\end{equation}
The change in entropy of the tape is given by \cite{Esposito2010} 
\begin{equation}
  \label{eq:DS_tape}
  \begin{split}
    \Delta S_\ta 
    =& -\sum_y p^\out_\ta(y)\ln p^\out_\ta(y) 
    + \sum_y p^\equ_\ta(y)\ln p^\equ_\ta(y)\\
    =& -\sum_y p^\out_\ta(y)\ln\frac{p^\out_\ta(y)}{p^\equ_\ta(y)}\\
    &+\frac{1}{T_\ta}\sum_y (p^\equ_\ta(y)-p^\out_\ta(y))(F-E(y))\\
    =& -\frac{Q_\ta}{T_\ta}-D[p^\out_\ta(\cdot)|p^\equ_\ta(\cdot)],
  \end{split}
\end{equation}
where $D[p(\cdot)|q(\cdot)] = \sum_x p(x) \ln p(x)/q(x)$ denotes the  Kullback--Leibler divergence of $p(x)$ from $q(x)$.

Because of the cyclic state of the device the first law of thermodynamics acquires the form
\begin{equation}\label{eq:1stlaw}
 W + Q_\ba + Q_\ta = 0.
\end{equation} 
The second law stipulates 
\begin{equation}\label{eq:2ndLaw}
 \Delta S_\ba +\Delta S_\ta \geq 0,
\end{equation} 
where the entropy change of the bath is given by $\Delta S_\ba = -{Q_\ba}/{T_\ba}$.
Using \eqref{eq:1stlaw}, \eqref{eq:2ndLaw} and \eqref{eq:DS_tape} we derive an upper bound for the efficiency of the device working as a heat engine:
\begin{equation}
  \label{eq:eta_max}
  \eta\equiv \frac{-W}{Q_\ba} \leq \eta_\mathrm{max}=1- \frac{T_\ta}{T_\ba} -\frac{T_\ta}{Q_\ba} D[p^\out_\ta(\cdot)|p^\equ_\ta(\cdot)].
\end{equation}
For the correct interpretation of this bound it is crucial to keep in mind the twofold nature of a thermal tape. Since the Kullback--Leibler divergence is always positive \cite[Chap.~2]{Cover2006},  the efficiency of the setup always stays {\em below} the Carnot efficiency $\eta_C = 1-T_\ta/T_\ba$. The Carnot limit is reached only if the cells relax infinitely fast to their equilibrium distribution. In such a reasoning the tape plays the role of a (non-ideal) heat bath and the presence of some slowly relaxing degrees of freedom memorizing the state of the system {\em prevents} the setup from reaching the limits allowed by the Second Law of thermodynamics \footnote{Note in particular the difference between \eqref{eq:eta_max} and Eq.~(54) in \cite{Deffner2013} in which an efficiency {\em larger} than the corresponding Carnot value may be found.}. In the ideal tape limit $E(y)=\epsilon(y)\, T_\ta$ with $T_\ta\to 0$, on the other hand, we find $Q_t \to 0$ from \eqref{eq:defQt} whereas \eqref{eq:DS_tape} shows that $\Delta S_t$ remains non-zero, i.e. the system still exchanges entropy with the tape but no energy. The bound \eqref{eq:eta_max} approaches its trivial limit $\eta\leq 1$ and the system is able now to perform {\em beyond} the restrictions imposed by the (traditional) Second Law since it periodically extracts energy from a heat bath and converts it into work (accompanied by writing information on the tape). In this interpretation of the tape as an information source \cite{Deffner2013} the very same correlations between system and tape that were detrimental for its role as a heat bath turn out to be the clue for its performance over the (traditional) Second Law. The result \eqref{eq:eta_max} holds for autonomous systems; it remains valid also for systems with measurement and feedback as long as no additional entropy is absorbed or generated during the measurement process. 


\section{Autonomous device}
\label{sec:example1}

In this section we detail the general analysis described above for a simple autonomous system with exactly solvable dynamics. The setup is motivated by a recently introduced model of a Maxwell's demon \cite{Mandal2012}, see also \cite{Barato2013a,Barato2013}. We assume that there are only two states per cell of the tape, $y=0,1$, with energies  $E(0)=0$ and $E(1) =E_\ta\equiv \epsilon T_\ta  >0 $. The restriction to two levels is merely made for formal simplicity; it has the additional advantage that any distribution of  incoming bits may be modeled by a tape temperature $T_\ta$. The scaling of $E_\ta$ with $T_\ta$ ensures that the ratio between up and down cells in the incoming tape is fixed to $\tanh(\epsilon/2)$ irrespective of the value of $T_\ta$.

The device is assumed to be a two-level system as well, with states denoted by $x=0,1$ and energies $E_\dev(0)=0$ and $E_\dev(1)\equiv E_\dev>0$. 

\begin{figure}
  \centering
  \includegraphics{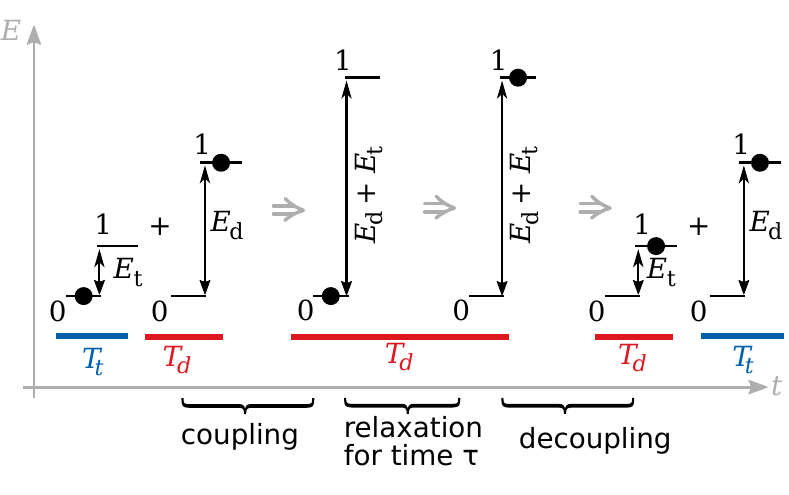}
  \caption{Energetics of the autonomous device. The two level systems of the tape and the device are coupled to form one two level system. The combined system relaxes for some time $\tau$ and is finally decoupled. }
  \label{fig:JM}
\end{figure}

The interaction between device and tape is defined as follows, cf. Fig.~\ref{fig:JM}. At the beginning of the interaction interval device and tape are instantaneously coupled to form a composite two-level system with energies $0$ and $E_\ta+E_\dev$. The combined system is put into the state of the tape. The energy required for this setting of the initial condition is obtained from the work source. For a time $\tau$ the combined system then relaxes in contact with the heat bath at temperature $T_\ba$ under the exchange of heat $Q_\ba$. At the end of the interaction period the coupling between device and tape is instantaneously removed and both remain in their respective states. The tape hence records the final state of the device. Neither heat nor work is exchanged in this final step. The tape is then moved one step and the device starts to interact with the next cell.

Denoting the probability of the device to be in state $x=0,1$ at time $t$ by $p_\dev(x,t)$, the stochastic evolution of the combined system during the interaction with the heat bath is governed by the master equation
\begin{equation}\label{eq:ME}
  \frac{\diff}{\diff t}
  \begin{pmatrix}
    p_\dev(0,t)\\
    p_\dev(1,t)
  \end{pmatrix}
=
 \begin{pmatrix}
    -W_{1,0}^\dev & W_{0,1}^\dev\\
    W_{1,0}^\dev & -W_{0,1}^\dev
  \end{pmatrix}
  \begin{pmatrix}
    p_\dev(0,t)\\
    p_\dev(1,t)
  \end{pmatrix},
\end{equation}
with transition rates
\begin{equation}
  \label{eq:reates_demon}
    W_{1,0}^\dev = \e^{-\frac{E_\dev+\epsilon T_\ta }{2 T_\dev}},\quad 
    W_{0,1}^\dev = \e^{+\frac{ E_\dev+ \epsilon T_\ta}{2 T_\dev}}.
\end{equation}
The solution of \eqref{eq:ME} for the initial condition $p_\dev(y,0) = p_\ta^\equ (y)$ is of the form
 \begin{equation}
  \label{eq:p_demon}
  p_\dev(y,t) = \e^{-r t} p_\ta^\equ(y) + (1-\e^{-r t})p_\dev^\equ(y),
\end{equation}
where 
\begin{equation}
   \label{eq:peq_demon}
    p_\dev^\equ(0) = \frac{W_{0,1}^\dev}{r},\quad
    p_\dev^\equ(1) = \frac{W_{1,0}^\dev}{r},
  \end{equation}
denotes the equilibrium distribution the system approaches for $\tau\to\infty$. Here $r\equiv W_{0,1}^\dev + W_{1,0}^\dev$. The probability to find the device in state $x$ at the end of the interaction interval is hence given by $p_d(x,\tau)$.

Defining
\begin{equation}
  \begin{split}
    \label{eq:Phi}
    \Phi(\tau) &:= [p_\ta^\equ(1) - p_\dev(1,\tau)] \\
    &= [p_\ta^\equ(1) - p_\dev^\equ(1)](1-\e^{-r\tau}).
  \end{split}
\end{equation}
we find for the average work $W$ provided by the work reservoir 
\begin{equation}
  \label{eq:W_ML}
  W = \Phi(\tau)  E_\dev,
\end{equation}
and for the average heat $Q_\ba$ exchanged between device and heat bath
\begin{equation}
  \label{eq:Q_ML}
  Q_\ba = -\Phi(\tau) (E_\dev+\epsilon  T_\ta).
\end{equation}
The efficiency is given by 
\begin{equation}
  \label{eq:eta_JM}
 \eta_\mathrm{aut} = \frac{-W}{Q_\ba} = 1- \frac{\epsilon T_\ta}{\epsilon T_\ta +E_\dev }. 
\end{equation}

The device delivers work, $W < 0$, if $\Phi(\tau) < 0$. With \eqref{eq:p_eq_tape} and \eqref{eq:peq_demon} this condition acquires the form
\begin{equation}
  \label{eq:condition}
  \frac{T_\ta}{T_\dev} < \frac{\epsilon T_\ta}{\epsilon T_\ta + E_\dev }.
\end{equation}
Similarly to \eqref{eq:eta_max} this relation implies, that the efficiency is below the Carnot  limit $\eta_\mathrm{C}$ whenever the device delivers work. 

\begin{figure}
  \centering
  \includegraphics{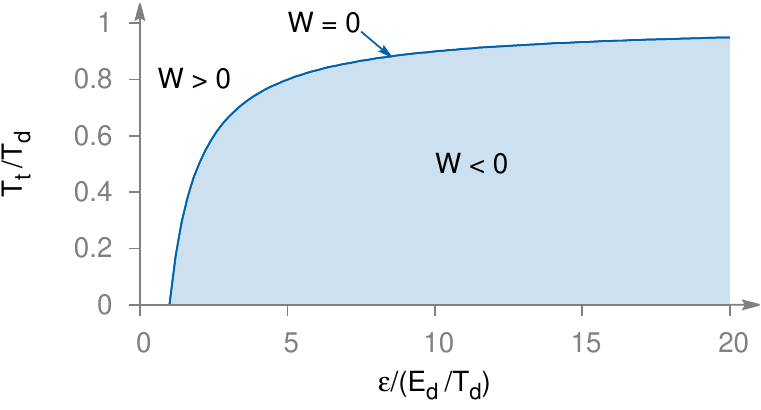}
  \caption{Regime of work production. If the parameters lie in the shaded region, the device delivers work. If they are in the white area, the device consumes work. On the solid line, dividing both regions, the system is always in equilibrium and the work is zero.}
  \label{fig:regime}
\end{figure}

The regime of work production is shown in Fig.~\ref{fig:regime}. For $\epsilon \to \infty$ the incoming tape is saturated with zeros and the device delivers work for all $T_\ta < T_\ba$. With decreasing $\epsilon$ the maximal value of $T_\ta$ at which work is still delivered gets smaller. Finally, if $\epsilon < E_\dev / T_\dev$ the device consumes work for all $T_\ta$. In the ideal tape limit, $T_\ta = 0$, work is produced for any  $\epsilon > E_\dev / T_\dev$. The line $W = 0$ is characterized by $p_\ta^\equ(x) = p_\dev^\equ(x)$ [cf. \eqref{eq:Phi}] and the system is in equilibrium during the entire process.

\begin{figure}
  \centering
  \includegraphics{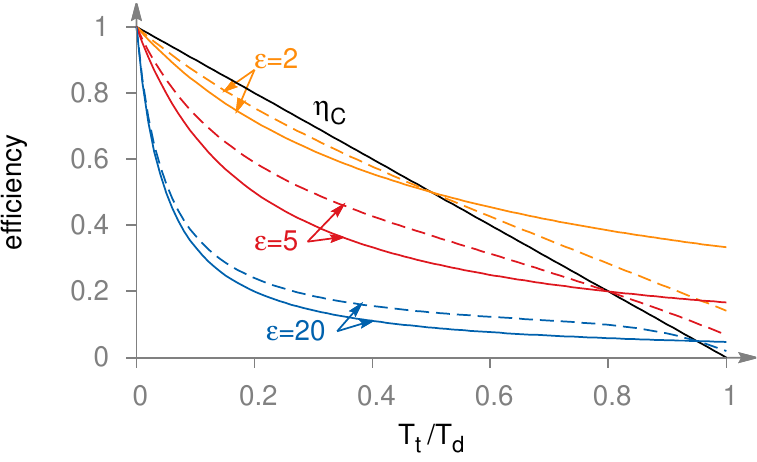}
  \caption{Comparison of the efficiency $\eta_\mathrm{aut}$ (solid lines) of the autonomous device with the upper bound $\eta_\mathrm{max}$ (dashed line) given by \eqref{eq:eta_max} and the Carnot efficiency $\eta_\mathrm{C}$  for $\tau = 10$ and $E_\dev = T_\dev = 1$.}
  \label{fig:efficiency}
\end{figure}

Fig.~\ref{fig:efficiency} compares the efficiency $\eta_\mathrm{aut}$ with the bound $\eta_\mathrm{max}$ and the Carnot value $\eta_\mathrm{C}$ for three different values of $\epsilon$. In the regime of work production we have $\eta_\mathrm{aut} \leq \eta_\mathrm{max} \leq \eta_\mathrm{C}$ as it should be. With the exact solution \eqref{eq:p_demon} of the dynamics at hand the complete entropy balance may be determined which demonstrates that the Second Law \eqref{eq:2ndLaw} is fulfilled as {\em in}equality whenever $W\neq 0$ . Accordingly, $\eta_\mathrm{aut}$ must be strictly smaller than $\eta_\mathrm{max}$. A saturation of the bounds $\eta_\mathrm{max}$ and $\eta_\mathrm{C}$ is obtained only for either the equilibrium situation, $W=0$, or in the ideal tape limit  $T_\ta = 0$. In the former case no work is performed since Carnot efficiency implies zero power which for a finite interval $\tau$ is equivalent to $W=0$. In the latter case all curves meet at $\eta=1$.


\section{System with feedback}
\label{sec:example2}
In this section we investigate a system driven by a protocol which depends on the outcome of a measurement of the system state. The result of the measurement is written on a tape and we are again interested in the role of the energy necessary to accomplish this storage. We use the same general setup as in the previous section with device and tape modeled as two-level systems. The incorporation of the feedback step is done in analogy to a model introduced in \cite{Barato2013}.

At the beginning of the interaction the states of the tape, $y_\inn$, and of the device,  $x_\inn$, are independent samples from the probability distributions $p_\ta^\equ(y_\inn)$ and $p_\dev^\inn(x_\inn)$, respectively. The state $x_\inn$ of the device is then measured and the result is stored as $y_\out$ on the tape. We would like to include imprecise measurements into the analysis and therefore allow for $x_\inn\neq y_\out$ with a certain probability. It is convenient to use the incoming bits of the tape as preassigned compilation of these measurement errors \cite{Barato2013}. The measurement is then prescribed by 
\begin{equation}
  \label{eq:measurement}
  y_\out =
  \begin{cases}
    x_\inn &\text{ if } y_\inn = 0,\\
    1- x_\inn &\text{ if } y_\inn = 1,\\
  \end{cases}
\end{equation}
and the probability for an erroneous measurement is \mbox{$1/(1+\e^\epsilon)$}. Eq.~\eqref{eq:measurement} defines a one-to-one correspondence between the state $(x_\inn,y_\inn)$ before and the state $(x_\inn,y_\out)$ after the measurement. As a result no entropy is generated during the measurement \cite{Landauer1961} and Eq.~\eqref{eq:eta_max} still holds. 

To extract work from the system the following feedback is performed. For $y_\out=0$, the system stays unchanged. For $y_\out=1$, i.e. when the system is more likely to be in the upper state, the energies of the states  $x=0$ and $x=1$ are instantly interchanged. If the system was indeed in the upper state, the energy $E_\dev$ is released as work. If not, the same amount of energy is consumed.  After this change of energies the labels of the states are interchanged as well, so that $x=1$ again denotes the state with higher energy. 

After this feedback step, the probability to find the device in the upper state is given by  $p_\ta^\equ(1)$, irrespective of $x_\inn$. The device now relaxes for the duration $\tau$ in contact with its heat bath. This relaxation is again governed by the master equation \eqref{eq:ME} where, however, the rates are now given by
\begin{equation}
  \label{eq:reates_feedback}
  \begin{split}
    W_{1,0}^\dev &= e^{-\frac{E_\dev }{2 T_\dev}},\\ 
    W_{0,1}^\dev &= e^{+\frac{ E_\dev}{2 T_\dev}}.
  \end{split}
\end{equation}
Keeping in mind these modified rates the ensuing analysis proceeds as in the previous section:
the solution $p_\dev(x,t)$ of the master equation is given by \eqref{eq:p_demon} with \eqref{eq:peq_demon} denoting the equilibrium distribution that the system approaches for large $\tau$. 

In the steady state of the system, the distribution of $x$ at the beginning and at the end of the interaction must be the same, $p_\dev^\inn(x) = p_\dev(x,\tau)$. Using 
\begin{equation}
  p_\ta^\out(1) 
  = p_\ta^\equ(0)\, p_\dev^\inn(1) + p_\ta^\equ(1)\, p_\dev^\inn(0)
\end{equation}
we find 
\begin{equation}
  \label{eq:Q_tape_feedback_exapmle}
  \begin{split}
    Q_\ta =& \left[ p_\ta^\equ(1) - p_\ta^\out(1) \right] \epsilon T_\ta \\
    =& 
    \left[
      p_\ta^\equ(1) - \Phi(\tau)
    \right]
    \left[
      p_\ta^\equ(1) - p_\ta^\equ(0) 
    \right] \epsilon T_\ta.
  \end{split}
\end{equation}
as well as
\begin{equation}
  \label{eq:Qb_feedback}
  Q_\ba = -\Phi(\tau) E_\dev.
\end{equation}
With $W=-(Q_\ba+Q_\ta)$ this yields for the efficiency 
\begin{equation}
  \label{eq:eta_feedback}
  \eta_\mathrm{fb} = \frac{-W}{Q_\ba} = 
  1 - \frac{
    \left[
      p_\ta^\equ(1) - \Phi(\tau)
    \right]
    \left[
      p_\ta^\equ(1) - p_\ta^\equ(0) 
    \right] \epsilon T_\ta
  }{
    \Phi(\tau) E_\dev
  }.
\end{equation}

\begin{figure}
  \centering
  \includegraphics{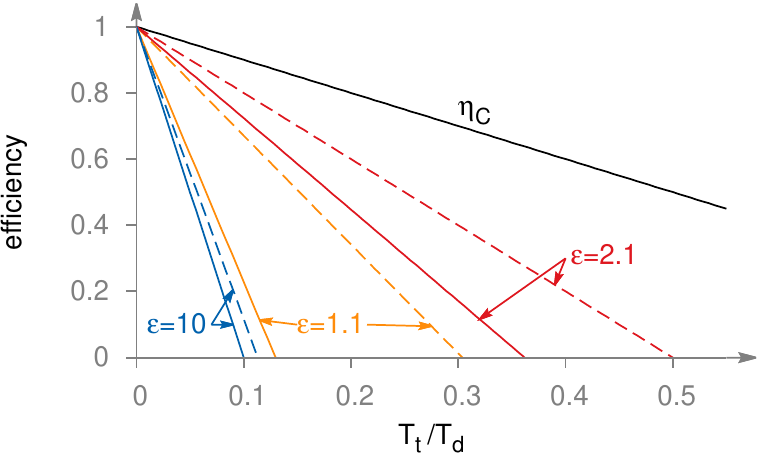}
  \caption{Comparison of the efficiency $\eta_\mathrm{fb}$ (solid lines) of the device with feedback control with $\eta_\mathrm{max}$ (dashed lines) and the Carnot efficiency $\eta_\mathrm{C}$ for different values of $\epsilon$. In this plot $E_\dev = T_\ba = 1$ and $\tau = 10$.}
  \label{fig:eta_feedback}
\end{figure}

Fig.~\ref{fig:eta_feedback} compares this result with $\eta_\mathrm{max}$ and  $\eta_\mathrm{C}$ for different values of $\epsilon$. We find again  
$\eta_\mathrm{fb} < \eta_\mathrm{max} < \eta_\mathrm{C}$ for all $T_\ta > 0$, and in the ideal tape limit, $T_\ta\to 0$, all efficiencies converge to 1. However, in contrast to the autonomous case $W$ and $Q_\ba$ no longer change sign at the same value of $T_\ta$. Therefore $\eta_\mathrm{fb}=0$ when $W=0$. Also, there is no equilibrium situation for the system with feedback. 

With $\Delta S_\ba = -Q_\ba/T_\ba = (W + Q_\ta)/T$,  the second law  \eqref{eq:2ndLaw} acquires for the feedback system the form  
\begin{equation}
  \label{eq:W_max1}
   \frac{-W}{T_\ba} \leq  \Delta S_\ta + \frac{Q_\ta}{T_\ba}.
\end{equation}
Due to our dual use of the tape as source of measurement errors and chronicle of measurement outcomes we have \cite[ Chap.~7]{Cover2006}
\begin{equation}
  \label{eq:I=S_feedback}
  \Delta S_\ta=I(x_\inn,y_\out),
\end{equation}
with
\begin{equation}
  \label{eq:I_def}
  I(x_\inn,y_\out) = -\sum_{x_\inn,y_\out} p(x_\inn,y_\out) 
  \ln \frac{p(x_\inn,y_\out)}{p_\dev^\inn(x_\inn)p_\ta^\out(y_\out)}
\end{equation}
denoting the mutual information between system and tape that is induced by the measurement. 
We may hence rewrite \eqref{eq:W_max1} as 
\begin{equation}
  \label{eq:W_I_relation}
   -W \leq  I(x_\inn,y_\out) T_\ba + Q_\ta . 
\end{equation}
In the ideal tape limit, $T_\ta\to 0$, we have $Q_\ta \to 0$ and \eqref{eq:W_I_relation} assumes the well-known form of the maximum work theorem for systems with feedback \cite{Sagawa2012,Sagawa2010, Abreu2012, Deffner2013, Vaikuntanathan2011, Granger2011}
 \begin{equation}
  \label{eq:W_I_SU}
   -W \leq  I(x_\inn,y_\out) T_\ba.
\end{equation}
The relation \eqref{eq:W_I_relation}  is also a consequence of the Sagawa-Ueda fluctuation theorem \cite{Sagawa2012a}. 
Since $Q_\ta\leq 0$ as follows from \eqref{eq:Q_tape_feedback_exapmle} we hence find that a tighter bound for the maximally extractable work in a feedback setup results if the energetic cost of information transfer is taken into account. 

\begin{figure}
  \centering
  \includegraphics{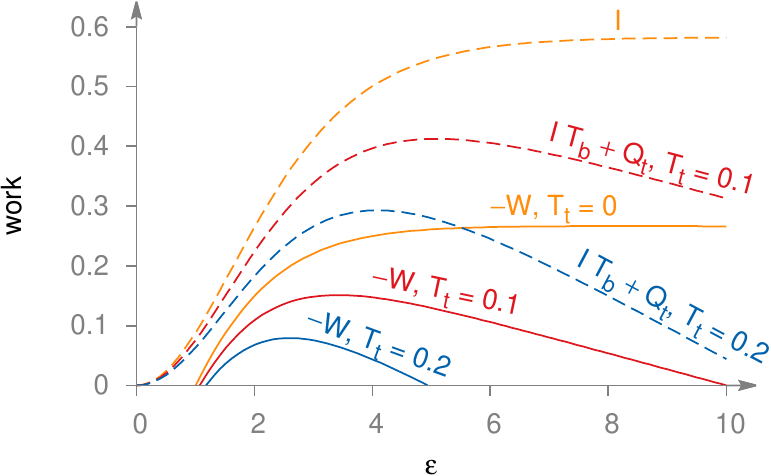}
  \caption{Work  $-W$ delivered by the feedback device (full lines) and upper bound given by \eqref{eq:W_I_relation} (dashed lines) as a function of the energy parameter $\epsilon$ of the tape for different tape temperatures $T_\ta$. The other parameters are $E_\dev = T_\ba = 1$ and $\tau = 10$.}
  \label{fig:feedback2}
\end{figure}

Fig.~\ref{fig:feedback2} displays $-W$ and the bound resulting from \eqref{eq:W_I_relation} as a function of $\epsilon$ for different values of $T_\ta$. In the ideal tape limit, $T_\ta=0$, both curves increase monotonously and saturate at finite values for large $\epsilon$, i.e. error-free measurements. For $T_\ta>0$ both $-W$ and the corresponding bound exhibit a maximum. This non-monotonous behaviour could already be identified in Fig.~\ref{fig:eta_feedback} with the efficiencies for $\epsilon =2.1$ being larger than those for both $\epsilon =1.1$ and $\epsilon =10$. It can be understood intuitively as follows: for small values of $\epsilon$ many erroneous measurements occur and therefore $-W$ as well as $I$ are small. For large $\epsilon$ measurement errors became rare but  the energy $-Q_\ta$ to write the measurement results to the tape grow. Accordingly, the difference between the bounds \eqref{eq:W_I_SU} and \eqref{eq:W_I_relation} increases and the latter must eventually go down again. Hence, if the energetic cost of information storage are properly accounted for the work production is not maximized for error-free measurements. Instead, there is an optimal fraction of errors.


\section{Summary}

\label{sec:conclusion}
Information is a thermodynamic resource; upon producing or consuming information physical systems may convert energy with efficiencies that are unfeasible otherwise. A direct but also  highly idealized way to study the thermodynamic impact of information consists in the sole inclusion of the Shannon entropy of the information source into the overall entropy balance of the system \cite{Andrieux2008,Andrieux2009,Barato2013,Barato2013a,Cao2009,Deffner2013, Mandal2012}.

However, information transfer is typically accompanied by energy transfer as well. Introducing a thermal tape as a succession of multi-stable cells with nontrivial intrinsic energetics we discussed some basic implications of the energetic cost in information exchange. The \emph{thermal tape} may be either interpreted as a (non-ideal) heat bath or as a (non-ideal) information source. The correlations between system and tape that are introduced by the interaction are crucial in both cases. In the former perspective they {\em reduce} the efficiency which therefore stays below the corresponding Carnot value. In the latter, they may {\em increase} the efficiency and performance beyond the limits set by the (traditional) Second Law of Thermodynamics becomes possible. 

We illustrated our general results with the detailed analysis of two exactly solvable model situations. The first is an autonomous stochastic system. Depending on the interpretation of the thermal tape it may act as a conventional heat engine or as a Maxwell's demon. The second systems is driven by measurement dependent feedback. Here we showed that the bound for the efficiency of work extraction becomes tighter when the energetic cost of information transfer is take into account, and that the efficiency shows a non-trivial maximum at a finite fraction of measurement errors. This is in contrast to the situation neglecting the energy exchange in information transfer where the efficiency gets maximal for zero measurement error. 

If we prescribe the same energy to all states of the tape we come back to the idealization in which no energy is necessary to write on the tape. In this {\em ideal tape limit} our findings  reduce to the results for efficiencies and maximum work limits known in the literature \cite{Sagawa2010,Sagawa2012, Abreu2012, Deffner2013, Vaikuntanathan2011, Granger2011}. There is hence a certain complementarity between the thermodynamic concepts of work and information: whereas work is energy without entropy, information is (in its idealized form) entropy without energy.

\begin{acknowledgements}
  It is a pleasure to thank Massimiliano Esposito for stimulating
  discussions. Financial support from DFG under EN278/9-1 is
  gratefully acknowledged.
\end{acknowledgements}

\bibliographystyle{apsrev}
\bibliography{ref-clean}

\begin{thebibliography}{21}
\expandafter\ifx\csname natexlab\endcsname\relax\def\natexlab#1{#1}\fi
\expandafter\ifx\csname bibnamefont\endcsname\relax
  \def\bibnamefont#1{#1}\fi
\expandafter\ifx\csname bibfnamefont\endcsname\relax
  \def\bibfnamefont#1{#1}\fi
\expandafter\ifx\csname citenamefont\endcsname\relax
  \def\citenamefont#1{#1}\fi
\expandafter\ifx\csname url\endcsname\relax
  \def\url#1{\texttt{#1}}\fi
\expandafter\ifx\csname urlprefix\endcsname\relax\def\urlprefix{URL }\fi
\providecommand{\bibinfo}[2]{#2}
\providecommand{\eprint}[2][]{\url{#2}}

\bibitem[{\citenamefont{Seifert}(2012)}]{Seifert2012}
\bibinfo{author}{\bibfnamefont{U.}~\bibnamefont{Seifert}},
  \bibinfo{journal}{Reports Prog. Phys.} \textbf{\bibinfo{volume}{75}},
  \bibinfo{pages}{126001} (\bibinfo{year}{2012}).

\bibitem[{\citenamefont{Esposito}(2012)}]{Esposito2012a}
\bibinfo{author}{\bibfnamefont{M.}~\bibnamefont{Esposito}},
  \bibinfo{journal}{Phys. Rev. E} \textbf{\bibinfo{volume}{85}},
  \bibinfo{pages}{041125} (\bibinfo{year}{2012}).

\bibitem[{\citenamefont{Cao and Feito}(2009)}]{Cao2009}
\bibinfo{author}{\bibfnamefont{F.}~\bibnamefont{Cao}} \bibnamefont{and}
  \bibinfo{author}{\bibfnamefont{M.}~\bibnamefont{Feito}},
  \bibinfo{journal}{Phys. Rev. E} \textbf{\bibinfo{volume}{79}},
  \bibinfo{pages}{041118} (\bibinfo{year}{2009}).

\bibitem[{\citenamefont{Sagawa and Ueda}(2010)}]{Sagawa2010}
\bibinfo{author}{\bibfnamefont{T.}~\bibnamefont{Sagawa}} \bibnamefont{and}
  \bibinfo{author}{\bibfnamefont{M.}~\bibnamefont{Ueda}},
  \bibinfo{journal}{Phys. Rev. Lett.} \textbf{\bibinfo{volume}{104}},
  \bibinfo{pages}{090602} (\bibinfo{year}{2010}).

\bibitem[{\citenamefont{Sagawa and Ueda}(2012{\natexlab{a}})}]{Sagawa2012a}
\bibinfo{author}{\bibfnamefont{T.}~\bibnamefont{Sagawa}} \bibnamefont{and}
  \bibinfo{author}{\bibfnamefont{M.}~\bibnamefont{Ueda}},
  \bibinfo{journal}{Phys. Rev. Lett.} \textbf{\bibinfo{volume}{109}},
  \bibinfo{pages}{180602} (\bibinfo{year}{2012}{\natexlab{a}}).

\bibitem[{\citenamefont{Deffner and Jarzynski}(2013)}]{Deffner2013}
\bibinfo{author}{\bibfnamefont{S.}~\bibnamefont{Deffner}} \bibnamefont{and}
  \bibinfo{author}{\bibfnamefont{C.}~\bibnamefont{Jarzynski}},
  \bibinfo{journal}{Phys. Rev. X} \textbf{\bibinfo{volume}{3}},
  \bibinfo{pages}{041003} (\bibinfo{year}{2013}).

\bibitem[{\citenamefont{Horowitz and Parrondo}(2011)}]{Horowitz2011a}
\bibinfo{author}{\bibfnamefont{J.~J.~M.} \bibnamefont{Horowitz}}
  \bibnamefont{and} \bibinfo{author}{\bibfnamefont{J.~M.~R.}
  \bibnamefont{Parrondo}}, \bibinfo{journal}{Europhys. Lett.}
  \textbf{\bibinfo{volume}{95}}, \bibinfo{pages}{10005} (\bibinfo{year}{2011}).

\bibitem[{\citenamefont{Granger and Kantz}(2011)}]{Granger2011}
\bibinfo{author}{\bibfnamefont{L.}~\bibnamefont{Granger}} \bibnamefont{and}
  \bibinfo{author}{\bibfnamefont{H.}~\bibnamefont{Kantz}},
  \bibinfo{journal}{Phys. Rev. E} \textbf{\bibinfo{volume}{84}},
  \bibinfo{pages}{061110} (\bibinfo{year}{2011}).

\bibitem[{\citenamefont{Abreu and Seifert}(2011{\natexlab{a}})}]{Abreu2012}
\bibinfo{author}{\bibfnamefont{D.}~\bibnamefont{Abreu}} \bibnamefont{and}
  \bibinfo{author}{\bibfnamefont{U.}~\bibnamefont{Seifert}},
  \bibinfo{journal}{Phys. Rev. Lett.} \textbf{\bibinfo{volume}{108}},
  \bibinfo{pages}{030601} (\bibinfo{year}{2011}{\natexlab{a}}).

\bibitem[{\citenamefont{Andrieux and Gaspard}(2008)}]{Andrieux2008}
\bibinfo{author}{\bibfnamefont{D.}~\bibnamefont{Andrieux}} \bibnamefont{and}
  \bibinfo{author}{\bibfnamefont{P.}~\bibnamefont{Gaspard}},
  \bibinfo{journal}{Proc. Natl. Acad. Sci. U. S. A.}
  \textbf{\bibinfo{volume}{105}}, \bibinfo{pages}{9516} (\bibinfo{year}{2008}).

\bibitem[{\citenamefont{Andrieux and Gaspard}(2009)}]{Andrieux2009}
\bibinfo{author}{\bibfnamefont{D.}~\bibnamefont{Andrieux}} \bibnamefont{and}
  \bibinfo{author}{\bibfnamefont{P.}~\bibnamefont{Gaspard}},
  \bibinfo{journal}{J. Chem. Phys.} \textbf{\bibinfo{volume}{130}},
  \bibinfo{pages}{014901} (\bibinfo{year}{2009}).

\bibitem[{\citenamefont{Barato and Seifert}(2013{\natexlab{a}})}]{Barato2013}
\bibinfo{author}{\bibfnamefont{a.~C.} \bibnamefont{Barato}} \bibnamefont{and}
  \bibinfo{author}{\bibfnamefont{U.}~\bibnamefont{Seifert}},
  \bibinfo{journal}{Europhys. Lett.} \textbf{\bibinfo{volume}{101}},
  \bibinfo{pages}{60001} (\bibinfo{year}{2013}{\natexlab{a}}).

\bibitem[{\citenamefont{Abreu and Seifert}(2011{\natexlab{b}})}]{Abreu2011}
\bibinfo{author}{\bibfnamefont{D.}~\bibnamefont{Abreu}} \bibnamefont{and}
  \bibinfo{author}{\bibfnamefont{U.}~\bibnamefont{Seifert}},
  \bibinfo{journal}{Europhys. Lett.} \textbf{\bibinfo{volume}{94}},
  \bibinfo{pages}{10001} (\bibinfo{year}{2011}{\natexlab{b}}).

\bibitem[{\citenamefont{Barato and Seifert}(2013{\natexlab{b}})}]{Barato2013a}
\bibinfo{author}{\bibfnamefont{A.~C.} \bibnamefont{Barato}} \bibnamefont{and}
  \bibinfo{author}{\bibfnamefont{U.}~\bibnamefont{Seifert}},
  \bibinfo{journal}{arXiv:1308.4598 [cond-mat.stat-mech]}
  (\bibinfo{year}{2013}{\natexlab{b}}).

\bibitem[{\citenamefont{Mandal and Jarzynski}(2012)}]{Mandal2012}
\bibinfo{author}{\bibfnamefont{D.}~\bibnamefont{Mandal}} \bibnamefont{and}
  \bibinfo{author}{\bibfnamefont{C.}~\bibnamefont{Jarzynski}},
  \bibinfo{journal}{Proc. Natl. Acad. Sci. U. S. A.}
  \textbf{\bibinfo{volume}{109}}, \bibinfo{pages}{11641}
  (\bibinfo{year}{2012}).

\bibitem[{\citenamefont{Vaikuntanathan and
  Jarzynski}(2011)}]{Vaikuntanathan2011}
\bibinfo{author}{\bibfnamefont{S.}~\bibnamefont{Vaikuntanathan}}
  \bibnamefont{and}
  \bibinfo{author}{\bibfnamefont{C.}~\bibnamefont{Jarzynski}},
  \bibinfo{journal}{Phys. Rev. E} \textbf{\bibinfo{volume}{83}},
  \bibinfo{pages}{061120} (\bibinfo{year}{2011}).

\bibitem[{\citenamefont{Horowitz et~al.}(2013)\citenamefont{Horowitz, Sagawa,
  and Parrondo}}]{Horowitz2013}
\bibinfo{author}{\bibfnamefont{J.~M.} \bibnamefont{Horowitz}},
  \bibinfo{author}{\bibfnamefont{T.}~\bibnamefont{Sagawa}}, \bibnamefont{and}
  \bibinfo{author}{\bibfnamefont{J.~M.~R.} \bibnamefont{Parrondo}},
  \bibinfo{journal}{Phys. Rev. Lett.} \textbf{\bibinfo{volume}{111}},
  \bibinfo{pages}{010602} (\bibinfo{year}{2013}).

\bibitem[{\citenamefont{Esposito et~al.}(2010)\citenamefont{Esposito,
  Lindenberg, and van~den Broeck}}]{Esposito2010}
\bibinfo{author}{\bibfnamefont{M.}~\bibnamefont{Esposito}},
  \bibinfo{author}{\bibfnamefont{K.}~\bibnamefont{Lindenberg}},
  \bibnamefont{and} \bibinfo{author}{\bibfnamefont{C.}~\bibnamefont{van~den
  Broeck}}, \bibinfo{journal}{New J. Phys.} \textbf{\bibinfo{volume}{12}},
  \bibinfo{pages}{013013} (\bibinfo{year}{2010}).

\bibitem[{\citenamefont{Cover and Thomas}(2006)}]{Cover2006}
\bibinfo{author}{\bibfnamefont{T.~M.} \bibnamefont{Cover}} \bibnamefont{and}
  \bibinfo{author}{\bibfnamefont{J.~A.} \bibnamefont{Thomas}},
  \emph{\bibinfo{title}{{Elements of Information Theory}}}
  (\bibinfo{publisher}{John Wiley \& Sons}, \bibinfo{year}{2006}).

\bibitem[{\citenamefont{Landauer}(1961)}]{Landauer1961}
\bibinfo{author}{\bibfnamefont{R.}~\bibnamefont{Landauer}},
  \bibinfo{journal}{IBM J. Res. Dev.} \textbf{\bibinfo{volume}{5}},
  \bibinfo{pages}{183} (\bibinfo{year}{1961}).

\bibitem[{\citenamefont{Sagawa and Ueda}(2012{\natexlab{b}})}]{Sagawa2012}
\bibinfo{author}{\bibfnamefont{T.}~\bibnamefont{Sagawa}} \bibnamefont{and}
  \bibinfo{author}{\bibfnamefont{M.}~\bibnamefont{Ueda}},
  \bibinfo{journal}{Phys. Rev. E} \textbf{\bibinfo{volume}{85}},
  \bibinfo{pages}{021104} (\bibinfo{year}{2012}{\natexlab{b}}).

\end{thebibliography}
\end{document}